\documentclass{iau}
\usepackage{graphicx}

\title[Galactic Dynamics feeding the Galactic Center]{Galactic Dynamics feeding the\\ Galactic Center}

\author[Florent Renaud, Eric Emsellem \& Fr\'ed\'eric Bournaud]{Florent Renaud$^1$, Eric Emsellem$^{2,3}$ \and Fr\'ed\'eric Bournaud$^1$}

\affiliation{$^1$Laboratoire AIM Paris-Saclay, CEA/IRFU/SAp, Universit\'e Paris Diderot, F-91191 Gif-sur-Yvette Cedex, France \\ email: {\tt florent.renaud@cea.fr} \\[\affilskip]
$^2$European Southern Observatory, 85748 Garching bei Muenchen, Germany\\
$^3$Universit\'e Lyon 1, Observatoire de Lyon, CRAL et ENS, 9 Av Charles Andr\'e, F-69230 Saint-Genis Laval, France}

\pubyear{2013}
\volume{303}  
\pagerange{}
\jname{The Galactic Center: Feeding and Feedback in a Normal Galactic Nucleus}
\editors{}
\begin{document}

\maketitle

\begin{abstract}
We present the hierarchical structure of the gas and its rapid evolution in the central region of a simulation of the entire Milky Way, run at subparsec resolution. We emphasize the coupling between the kpc-scale dynamics, the molecular ring and the central 5 pc disk feeding the super massive black hole.
\end{abstract}


The interplay between a fast evolving stellar background, inner resonances and gas dynamics makes the central kpc of every disk galaxy, and of the Milky Way in particular, a complex and yet not fully understood environment. In particular, the gas fueling toward the central black hole depends on the galactic dynamics at large scale. Such inflow rate appears to be a complex function of position, since several mechanisms and coupling between them can accelerate the gas flow or stop it at certain radii (e.g. resonances). The rapid evolution of this region modifies the dynamics locally, which eventually changes the properties of the inflow. Understanding the fine details of the center of a given galaxy requires an accurate knowledge of many parameters like the mass profile, the velocity distribution etc. Such information is barely accesible from observations of external galaxies because of limited resolution, and studies in the Milky Way suffer from projection and extinction effects. We use a hydrodynamical simulation to address these issues and monitor the evolution of the structures in the galactic center.

\section{Simulation}

Our analysis is based on a self-consistent hydrodynamical simulation of a Milky Way-like galaxy, run with the adaptive mesh refinment code RAMSES \cite[(Teyssier, 2002)]{Teyssier2002} and presented in details in \cite[Renaud et al. (2013)]{Renaud2013b}. Dark matter and stellar dynamics provide a ``live'' background for the gas component that we resolve down to 0.05 pc. Star formation and stellar feedback comprising photoionisation of H{\sc ii} regions, radiative pressure and supernova kinetic blasts are included. This short contribution presents the evolution of the gas structure. A much more detailed analysis of these structures, their interplay and the physics of the Galactic center will be soon presented in Emsellem et al. (in prep).

\section{Hierarchical structure}

\begin{figure}[b]
\begin{center}
\includegraphics[width=3.4in]{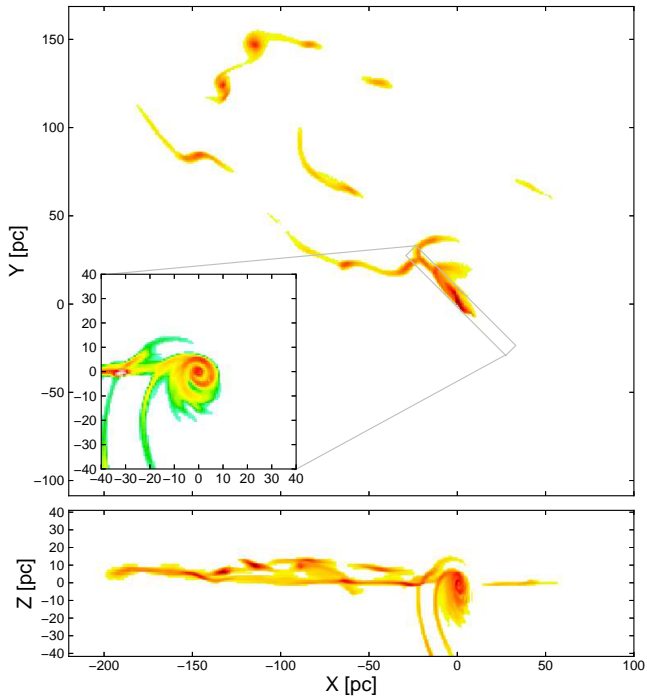} 
\caption{Volume density map of the gas denser than $10^4 \textrm{ cm}^{-3}$, in the galactic plane (top) and seen edge-on (bottom and thumbnail). At the instant shown here, the gaseous rings at 100 pc and 40 pc have already fragmented into dense star forming clumps. The clumps then fuel the central $\sim 10$ pc. Stellar feedback ejects some gas from the galactic plane, which creates the disk within 5 pc and arms perpendicular to the galactic disk when raining back onto it.}
\label{f1}
\end{center}
\end{figure}

Spiral arms and a bar form during the simulation within a couple of rotation periods, and exert torques on the gas which then fuels the central region. Mini-spirals convey the gas toward the central 100 pc. Because of galactic rotation, these spirals wrap into an elliptical ring-like structure in which the gas accumulates at a rate of $\sim 0.02 \textrm{ M}_\odot/\textrm{yr}$. Most of star formation in the central 100 pc takes place in this ring. As the net inflow rate eventually overcomes the local star formation rate, the gas mass in the ring increases (up to $3 \times 10^6 \textrm{ M}_\odot$). At the same time, secondary, smaller spirals convey gas toward the innermost Lindblad resonance at 40 pc (see \cite[Renaud et al., 2013]{Renaud2013b}, their Fig.\,8). When enough gas has been accumulated, the rings become unstable and fragment into dense clumps (Fig.\,\ref{f1}), which then migrate inward while forming star clusters. The gaseous left-overs and the young stars gather in the vicinity of the black hole, at about 5 pc.

During this process, stellar feedback strongly affects the structure of the ISM around the young stars formed in the rings and in clumps. Supernova blasts expanding inside the previously formed H{\sc ii} regions creates cavities of low gas density and eject some material away from the plane galactic disk (up to 150 pc), in the form of bubbles and dense filaments. When falling back onto the disk, close to the black hole, this gas forms a 20 pc wide disk, perpendicular to the galactic plane (Fig.\,\ref{f1}, bottom panel and thumbnail), which further accretes gas from the large scale fueling pictured above. Subsequent star formation takes place in this new disk, creating a complex distribution in the stellar population of the galactic center.

\end{document}